\documentclass[11pt, a4paper]{article}

\usepackage[utf8]{inputenc}
\usepackage{amsmath, amssymb, amsfonts} 
\usepackage{graphicx}                    
\usepackage{geometry}                    
\usepackage{hyperref}                    
\usepackage{natbib}                      
\usepackage{float}                       
\usepackage{booktabs}                    

\title{\textbf{Cooperation, Defection, and the Commons: \\ Reproducing an Agent-Based Model of Aksum with Three Logistic Equations}}
\author{Riccardo Vasellini$^{1,2}$, Federico Cecconi$^{1}$, Daniele Vilone$^{1,3}$ \\
\ \ \ \\
$^1$LABSS-ISTC, National Research Council, Rome, Italy; \\
$^2$BluEye Simulations, Grosseto, Italy; \\
$^3$GISC-Dep.to de Matem\'aticas, Universidad Carlos III de Madrid, Spain 
}  

\date{September 2026}

\begin{document}

\maketitle

\begin{abstract}
Cooperative behavior can improve environmental quality, whereas selfish defection can degrade shared resources. In a recent agent-based study of the Aksum civilization (Vasellini et al., 2025), a population of agents moving over a reconstructed landscape and playing a local Public Goods Game grew faster and peaked higher when the game was enabled, but settled, in the long run, at the same population level as when it was not. Here we show that this behavior does not depend on the many rules of that model. We develop a theoretical ecological model of three coupled discrete-time equations, in which two sub-populations grow logistically and share a carrying capacity that rises when cooperators are sufficiently numerous and decays otherwise. Without fitting any parameter to the simulation, the equations reproduce its qualitative behavior: an initial boom, damped oscillations, and stabilization; a higher peak when the cooperative mechanism is active; and a final equilibrium that does not depend on it. Cooperation can be self-sustained for a transient period, but in the absence of measures that mitigate free-riding it collapses, and the environmental gains it produced are lost with it.

\vspace{6pt}
\end{abstract}

\section{Introduction}
The ``Tragedy of the Commons'' suggests that individual rationality often leads to collective ruin in shared resource systems~\citep{hardin1968}, although a large body of empirical work has since shown that communities frequently avoid this outcome through institutions of their own making~\citep{ostrom1990}. Traditional ecological models often assume a static carrying capacity ($K$), ignoring the ability of organisms, in this case humans, to engineer their environment. This paper presents a mathematical model where $K$ is endogenous, evolving based on the ratio of cooperators ($A_t$) to defectors ($B_t$).

This work serves as the theoretical companion to the agent-based research presented by~\citet{vasellini2025} regarding the Aksum civilization. That model places agents on a landscape reconstructed from archaeological and GIS data, gives them vision, movement, metabolism, resource-based reproduction and strategy inheritance, and lets them play a spatially localized Public Goods Game (PGG) whose success raises the fertility of the surrounding land. It involves some twenty interacting rules. Its macroscopic outcome, however, is simple: the population grows, oscillates, and stabilizes; when the PGG is enabled the peak is higher; and the final population is the same in both cases. The aim of the present paper is to show that this outcome is reproduced by three coupled difference equations, and to explain, in those terms, why it occurs. We stress that no parameter of the equations is fitted to the simulation. The comparison is qualitative, and the claim is that the two models share the same structure but not the same numbers.

The idea that cooperation and the environment feed back on each other is not new. Population density has been coupled to public goods games in the ecological framework of~\citet{hauert2006}; a carrying capacity that grows with the fraction of cooperators appears in~\citet{melbinger2010}; and the general theory of games with environmental feedback has been developed by~\citet{weitz2016} and~\citet{tilman2020}. Under the name of niche construction, the same mechanism has been studied in evolutionary ecology~\citep{odlingsmee2003, krakauer2009}, and thresholds of the kind we use below are a standard ingredient of collective-risk dilemmas~\citep{santos2011}. Our contribution is not the feedback itself but its use as a reduction: we take a published, rule-rich, spatial simulation and show which small set of ingredients carries its macroscopic behavior.

The paper is organized as follows. Section~\ref{sec:abm} summarizes the agent-based model and what it does. Section~\ref{sec:model} presents the three equations. Section~\ref{sec:numerics} explores them numerically, Section~\ref{sec:comparison} compares them with the simulation, and Section~\ref{sec:analytical} gives an analytical account of the maximum of the cooperator population. We close with a discussion of what the reduction does and does not capture.

\section{The agent-based model}
\label{sec:abm}

The simulation of~\citet{vasellini2025} is built on the Sugarscape framework of~\citet{epstein1996}, in which agents collect a resource (``sugar'') from the patches they occupy, consume a fixed amount per time step, and die when their resources run out or when they reach a maximum age. To this core the Aksum model adds an empirical layer and a theoretical one. The empirical layer places the agents on a map of the Aksum area: patch fertility is seeded from a soil-productivity raster, rivers are imported from shapefiles, and the initial settlements are located at 698 pre- and proto-Aksumite sites known from survey. The theoretical layer is the set of rules that govern cooperation, summarized below.

\begin{itemize}
    \item \textbf{Movement and vision.} Agents scan patches within their vision range and move toward the most fertile ones, avoiding those already occupied.
    \item \textbf{Metabolism and reproduction.} Agents consume two units of sugar per step and may reproduce, at most twice, once they have stored twice their metabolic cost; offspring receive half the parent's resources.
    \item \textbf{Strategies.} Each agent is a full cooperator or a full defector. Offspring inherit the parent's strategy with a $5\%$ probability of mutation; all other traits are fixed.
    \item \textbf{The Public Goods Game.} Games are held at the proto-Aksumite sites, among the agents within a radius of three patches. Cooperators donate $10\%$ of their sugar; defectors donate nothing. The game succeeds if the total donation reaches a threshold $T = \alpha N + \beta R$, where $N$ is the number of participants, $R$ their total resources, and $\alpha = \beta = 0.05$.
    \item \textbf{Environmental feedback.} A successful game raises the maximum fertility of every patch within the radius by a fixed increment ($\Delta_f = 1.5$), up to a global cap. Independently of the game, every patch whose fertility exceeds its baseline loses a fixed amount ($\gamma = 0.7$) per step. Cooperation can therefore raise fertility temporarily, but the gain decays unless collective action is sustained.
\end{itemize}

The model was run in two scenarios, with the PGG enabled and with it disabled, for 100 independent realizations each, and the two population trajectories were compared step by step with a Welch $t$-test under a Bonferroni correction. Three findings are relevant here.

First, both scenarios display the same broad pattern: an initial exponential growth, followed by oscillations, and eventual stabilization. The oscillations, present in both runs, are characteristic of Sugarscape-type models, in which agents consume local resources, reproduce, and then experience crashes as resource availability falls below metabolic demand.

Second, the two scenarios differ in amplitude and peak but not in outcome. With the PGG the population grows faster and peaks higher, at around 3,200 agents against 2,400 without it; the difference is statistically significant during this early phase. As the population increases, however, donation becomes costlier, fewer agents contribute, fewer games succeed, the fertility boost decays, and the population converges back to the same equilibrium of roughly 2,100 agents reached in the non-cooperative scenario. At late times the difference between the two conditions is no longer significant.

Third, the same structure survives wide parameter changes. Raising the donation rate from $0.1$ to $0.5$ raises the peak substantially but leaves the long-run equilibrium within a few percent of 2,100; slowing the degradation of fertility raises the peak to around 4,000 agents, yet all curves converge on the same equilibrium. Across every lever tested, the sequence is the same: an early boom, damped oscillations, and a common final level.

This is the behavior the equations of the next section are meant to reproduce.

\section{Mathematical Formulation}
\label{sec:model}

We consider a system of two interacting sub-populations sharing a common resource pool. The state of the system at discrete time $t$ is defined by the population of cooperators $A_t$, the population of defectors $B_t$, and the dynamic carrying capacity of the environment $K_t$. Throughout, populations and capacities are measured in units of the baseline capacity $K_{min}$ defined below, so that all quantities plotted are dimensionless.

\subsection{Population Dynamics}
The populations are updated using coupled discrete-time logistic maps. To address the biological impossibility of negative populations in discrete models when resources are overshot, we apply a non-negativity constraint.

The evolution of cooperators ($A$) and defectors ($B$) is given by:

\begin{equation}
    A_{t+1} = \max\left( 0, \, r_A A_t \left( 1 - \frac{A_t + B_t}{K_t} \right) \right)
    \label{At_eq}
\end{equation}

\begin{equation}
    B_{t+1} = \max\left( 0, \, r_B B_t \left( 1 - \frac{A_t + B_t}{K_t} \right) \right)
    \label{Bt_eq}
\end{equation}

Where:
\begin{itemize}
    \item $N_t = A_t + B_t$ represents the total population density at time $t$.
    \item $r_A$ and $r_B$ represent the intrinsic growth rates. We assume $r_B \ge r_A$, reflecting the cost of cooperation (e.g., the donation cost in a Public Goods Game), which gives defectors a reproductive advantage. This assumption is fundamental to the definition of a social dilemma: cooperation entails a metabolic or opportunity cost (e.g., labor spent on land maintenance) that defectors avoid. Thus, in the absence of punishment or incentives to cooperation, free-riders possess a reproductive advantage~\citep{nowak2006}.
    \item The term $(1 - \frac{N_t}{K_t})$ represents density-dependent feedback modeled specifically as scramble competition~\citep{nicholson1954}, in the overcompensating form standard for discrete-time models~\citep{hassell1975}. Unlike contest competition, where resources are monopolized by dominant individuals allowing the population to stabilize at $K$, scramble competition implies an egalitarian distribution of resources. In this regime, when the total population $N_t$ significantly exceeds the carrying capacity $K_t$, resources are subdivided to the point where they become insufficient for the maintenance of any individual. Mathematically, this creates a strong non-linear overcompensation response where an overshoot at time $t$ leads to a sharp population crash at time $t+1$, driving the `Bust' phase observed in our results.
\end{itemize}

\subsection{Dynamic Carrying Capacity}
Unlike standard ecological models where $K$ is constant, we model the carrying capacity as an endogenous variable driven by social behavior. This models the mechanism where a ``successful'' Public Goods Game (PGG) increases the fertility of the land.

The update rule for $K_{t+1}$ is:

\begin{equation}
\label{Kdyn}
    K_{t+1} = \max\left( K_{min}, \, K_t - \delta_K + \Delta K \cdot \Theta(R_t - \theta) \right)
\end{equation}

Where:
\begin{itemize}
    \item $R_t = \frac{A_t}{N_t}$ is the cooperation ratio (proportion of cooperators).
    \item $K_{min}$ is the baseline carrying capacity (subsistence level) in the absence of enhancement.
    \item $\delta_K$ is the natural degradation rate of the environment per time step. This represents environmental entropy (e.g., soil erosion, irrigation decay) which naturally lowers capacity over time unless actively counteracted by cooperative maintenance.
    \item $\Delta K$ is the ``cooperation bonus,'' representing the increase of resources due to collective action.
    \item $\Theta(x)$ is the Heaviside step function:
    \begin{equation}
        \Theta(x) = \begin{cases}
        1 & \text{if } x \ge 0 \\
        0 & \text{if } x < 0
        \end{cases}
    \end{equation}
    \item $\theta$ is the cooperation threshold for which the PGG is considered successful.
\end{itemize}

\subsection{Relation to the agent-based model}
\label{sec:relation}
Each term of Eqs.~\eqref{At_eq}--\eqref{Kdyn} stands for a group of rules in the simulation. The two growth rates stand for reproduction from stored resources, with $r_A < r_B$ standing for the sugar that cooperators give away. The density term stands for competition for patches and for the crashes that follow overshoot. The threshold $\Theta(R_t - \theta)$ stands for the success condition of the PGG, which in the simulation depends on the number and resources of the participants but, as long as resources per agent do not vary too much, amounts to requiring that enough of them cooperate. The bonus $\Delta K$ and the decay $\delta_K$ stand for the fertility increment $\Delta_f$ and the degradation decrement $\gamma$ applied to the patches.

Just as important is what the equations leave out. There is no space, no movement and no vision; no individual resource stocks and no metabolism; no mutation of strategies; no ceiling on fertility; and the PGG is played by the whole population at once rather than locally around each site. Parameter values are listed in Table~\ref{tab:params}. They were chosen to display the phenomena clearly and are not calibrated to the simulation; the comparison of Section~\ref{sec:comparison} concerns the shape of the dynamics, not its numerical values.

\begin{table}[H]
\centering
\caption{Parameter values used in the figures, in units of $K_{min}$. Departures from these values are stated in the captions.}
\label{tab:params}
\begin{tabular}{lll}
\toprule
Symbol & Value & Meaning \\
\midrule
$K_{min}$   & $1$                   & baseline carrying capacity (unit) \\
$K_0$       & $1$                   & initial carrying capacity \\
$\delta_K$  & $1.5\times10^{-3}$    & degradation per time step \\
$\Delta K$  & $4.83\times10^{-3}$   & cooperation bonus per time step \\
$\theta$    & $0.5$                 & PGG success threshold on $R_t$ \\
$r_B$       & $2.995$               & defector growth rate \\
$r_A$       & $2.860$               & cooperator growth rate ($r_A/r_B = 0.955$) \\
$A_0$       & $0.020$               & initial cooperators \\
$B_0$       & $3.18\times10^{-4}$   & initial defectors ($R_0 = 0.98$) \\
\bottomrule
\end{tabular}
\end{table}

\section{Numerical exploration}
\label{sec:numerics}

We analyze the system behavior by varying the growth rates and the cost of cooperation.

\subsection{Growth Rate Sensitivity}
\begin{figure}[H]
    \centering
    \includegraphics[width=0.95\textwidth]{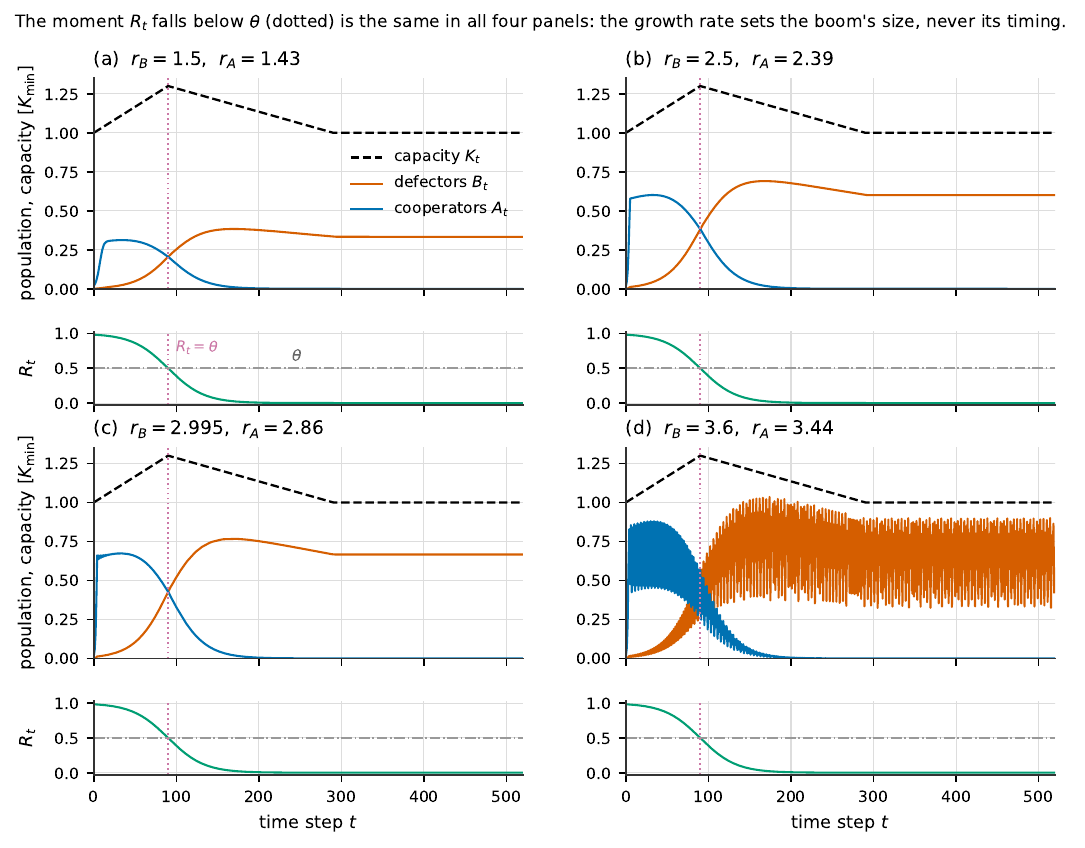}
    \caption{\textbf{Temporal Dynamics across Growth Regimes.} Cooperators ($A_t$, blue), defectors ($B_t$, orange) and carrying capacity ($K_t$, black dashed) for four values of $r_B$ at fixed $r_A/r_B = 0.955$; the strip below each panel shows the cooperation ratio $R_t$ against the threshold $\theta$ (grey, dash-dotted). The dotted vertical line marks the step at which $R_t$ falls below $\theta$ and the cooperation bonus switches off. Raising the growth rate changes the approach to equilibrium, from monotone in (a) to oscillatory in (b) and (c) and sustained oscillation in (d), and changes the size of the boom; it does not change the trajectory of the carrying capacity, which is identical in all four panels.}
    \label{fig:timeseries}
\end{figure}

Standard logistic maps exhibit period-doubling and chaos as $r$ increases beyond $3$~\citep{may1976}, and for $r > 4$ the quadratic map overshoots into negative values, which the non-negativity constraint in Eqs.~\eqref{At_eq}--\eqref{Bt_eq} interprets as an extinction event. (Alternative forms of density dependence, such as the exponential map of~\citet{ricker1954}, remain positive by construction and avoid this artifact; we retain the quadratic form for its transparency.) Our coupled model inherits this behavior. Figure~\ref{fig:timeseries} shows the temporal dynamics for four values of the defector growth rate, at a fixed cost of cooperation. At low growth rates the population approaches its equilibrium monotonically; at intermediate rates it approaches it through damped oscillations; at high rates the oscillations are sustained.

One feature of Figure~\ref{fig:timeseries} deserves emphasis. The carrying capacity follows exactly the same path in all four panels, and the cooperation ratio crosses the threshold at exactly the same step, although the four populations behave very differently. The reason is visible in Eqs.~\eqref{At_eq}--\eqref{Bt_eq}: the density factor $(1 - N_t/K_t)$ multiplies cooperators and defectors alike, so whatever the environment does to the population, it does equally to both, and the proportion between them changes only through the difference in growth rates. The decline of the cooperation ratio is therefore set by the cost of cooperation and the initial mix alone. The environment can make the boom larger; it cannot make it last longer.

\subsection{Bifurcation and sensitivity}
\begin{figure}[H]
    \centering
    \includegraphics[width=0.95\textwidth]{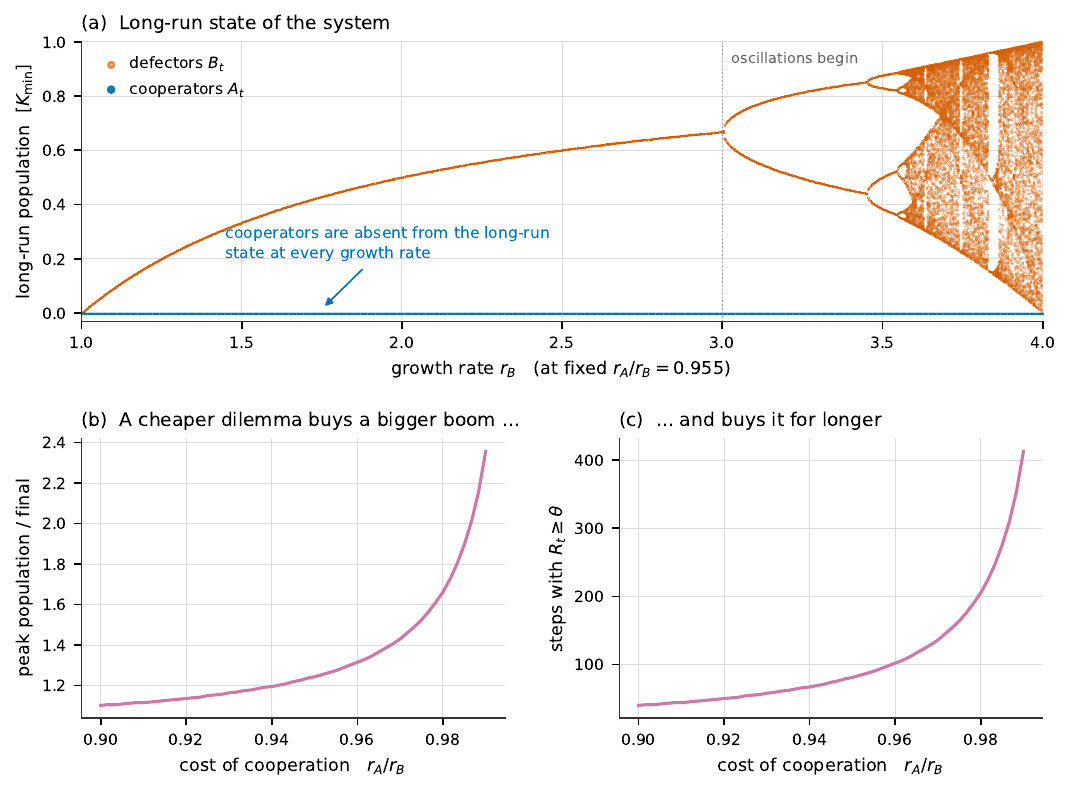}
    \caption{\textbf{Bifurcation and sensitivity.} \textbf{(a)} Long-run population of
    cooperators (blue) and defectors (orange) as a function of the growth rate $r_B$, at fixed
    $r_A/r_B = 0.955$, obtained by discarding the first 1280 of 1400 iterations. The defector
    branch follows the period-doubling route to chaos of the logistic map; the cooperator
    branch is zero throughout, because the cooperative phase is a transient and leaves no
    trace in the final state. \textbf{(b)} Peak population relative to the final equilibrium
    and \textbf{(c)} number of steps for which the cooperation bonus remains active, both as
    functions of the cost of cooperation $r_A/r_B$.}
    \label{fig:bifurcation}
\end{figure}

Figure~\ref{fig:bifurcation}(a) shows the long-run state of the system as the growth rate is
varied, at a fixed cost of cooperation. Two things are visible. The first is the familiar
period-doubling route to chaos of the logistic map, which the defector population inherits:
below $r_B = 3$ the system settles on a single level, above it the level splits in two, and
beyond $r_B \simeq 3.57$ the long-run behavior becomes irregular. The second is that the
cooperator branch is identically zero for every growth rate. This is not an artifact of the
range plotted: since defectors always grow faster and nothing compensates cooperators, the
cooperation ratio always falls to zero, the bonus in Eq.~\eqref{Kdyn} always switches off, and
the system always ends in the defector-only state supported by $K_{min}$. The cooperative
phase is a transient, and it does not appear in a diagram of the final state.

It is worth noting a consequence of the same reading. The population as a whole grows at a
rate lying between $r_A$ and $r_B$, weighted by the composition: close to $r_A$ while
cooperators are the majority, equal to $r_B$ once they are gone. Where the two rates straddle
$r_B = 3$, the loss of cooperators therefore moves the population from the settled side of
Figure~\ref{fig:bifurcation}(a) to the oscillating side. In that regime, the slower breeding
of cooperators, which is the cost they pay, is also what keeps the population steady.

Panels (b) and (c) show how the cooperative phase depends on that cost. The cheaper
cooperation is, the higher the population rises above its final level and the longer the
bonus remains active; both quantities grow without bound as the cost vanishes and the two
growth rates coincide. 

\subsection{The initial overshoot}
The non-negativity constraint has one further consequence. If the initial population is large compared with the initial capacity, the density factor is negative at the first step and both sub-populations are set to zero at once: the system collapses before cooperation has any chance to act. If instead the initial population is small, it survives long enough for the cooperation ratio to remain above $\theta$ while the capacity grows, and the boom follows. The boundary between the two regimes depends on $N_0/K_0$ and on the growth rates, not on the parameters of the environmental feedback.

\section{Comparison with the agent-based model}
\label{sec:comparison}

The dynamics of our difference equations are consistent with the agent-based simulations of the Aksum civilization conducted by~\citet{vasellini2025}. We stress again that the comparison is qualitative: the equations use the illustrative parameters of Table~\ref{tab:params}, and the question is whether they produce the same kind of behavior, not the same numbers.

\subsection{The Boom and Bust Cycle}
Our model produces a specific temporal pattern:
\begin{enumerate}
    \item \textbf{Boom:} When cooperators dominate ($R_t > \theta$), $K_t$ rises, allowing population growth.
    \item \textbf{Invasion:} Defectors multiply faster ($r_B > r_A$), reducing $R_t$.
    \item \textbf{Bust:} Once $R_t < \theta$, the environmental boost $\Delta K$ is lost. $K_t$ degrades via $\delta_K$, and the population returns to the level supported by the baseline $K_{min}$.
\end{enumerate}

This is the pattern of the simulation, in which cooperation ``can emerge spontaneously and temporarily raise collective outcomes, but in the absence of additional mechanisms it tends to break down''~\citep{vasellini2025}. The three phases reported there, ramp-up, decline and late equilibrium, correspond to the three steps above.

\subsection{The contrast between the two scenarios}
\begin{figure}[H]
    \centering
    \includegraphics[width=0.95\textwidth]{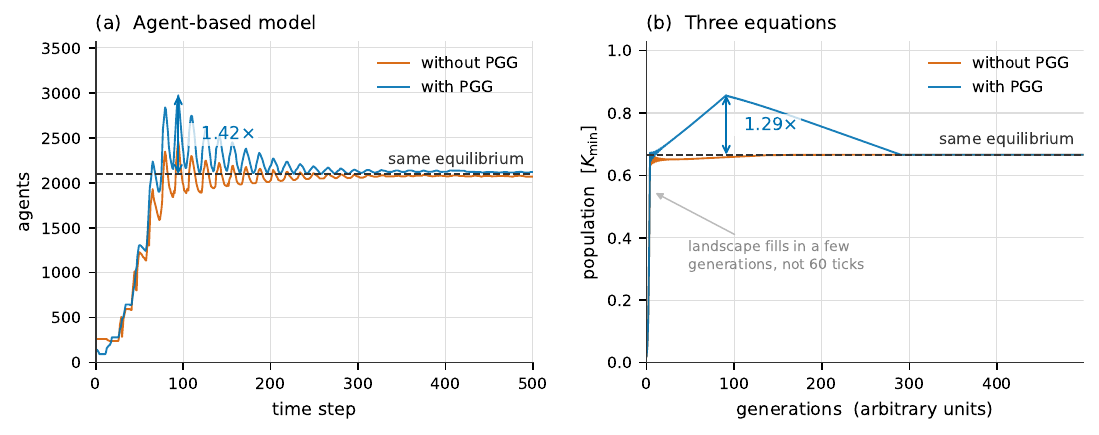}
    \caption{\textbf{The same contrast in the agent-based model and in the equations.} \textbf{(a)} Mean population of the Aksum simulation over 100 runs, with and without the Public Goods Game. \textbf{(b)} Total population from Eqs.~\eqref{At_eq}--\eqref{Kdyn} with the parameters of Table~\ref{tab:params}, and with $\Delta K = 0$ for the scenario without PGG. In both models the cooperative mechanism raises the peak (by a factor 1.42 in the simulation, 1.29 in the equations) and leaves the final equilibrium unchanged. The horizontal scales are not comparable: one iteration of the equations is a generation, whereas an agent in the simulation must accumulate resources over many steps before reproducing, so the landscape fills over tens of ticks there and within a few iterations here.}
    \label{fig:comparison}
\end{figure}

The most informative comparison is not between one simulated curve and one computed curve, but between the two scenarios of the simulation and the corresponding two scenarios of the equations. Figure~\ref{fig:comparison} shows them side by side. In the simulation the PGG is enabled or disabled; in the equations we set $\Delta K = 0$ for the second scenario and change nothing else.

The two panels share the same structure. In both, the cooperative mechanism raises the population above what the environment alone would support, and in both the population then falls back to exactly the level it reaches when the mechanism is absent. In the simulation this equality is an observation, reported three times: between the two scenarios, across the donation-rate sweep, and across the degradation-rate sweep. In the equations it is a consequence of their form. Since defectors grow faster, the cooperation ratio falls, the threshold in Eq.~\eqref{Kdyn} eventually fails, the capacity degrades back to $K_{min}$, and the population settles at the equilibrium of a logistic map with capacity $K_{min}$ and growth rate $r_B$. Nothing in that equilibrium depends on $\Delta K$, on $\delta_K$, on $\theta$, or on the initial composition. Whatever the cooperators built, the land gives back once they are gone.

The same reasoning explains why the difference between the two scenarios is significant only during the transient, as the simulation's step-by-step test shows: the two trajectories of the equations coincide before the boom, diverge while $K_t > K_{min}$, and coincide again afterwards.

\subsection{Sensitivity}
The sensitivity analysis of the simulation is consistent with the equations as well. A larger donation rate raises the peak but not the equilibrium; in the equations, a larger $\Delta K$ raises the peak capacity and leaves $K_{min}$ untouched. A slower degradation of fertility raises the peak; in the equations, a smaller $\delta_K$ does the same. In neither model does any environmental parameter move the final level.

\subsection{What the equations do not reproduce}
Table~\ref{tab:correspondence} lists the macroscopic features of the simulation and states, for each, whether the equations reproduce it. The slow initial rise of the simulated population is a matter of agents diffusing across a landscape and accumulating resources before reproducing; the equations have no space and reproduce every generation. The large, long-lived oscillations of the simulation arise from the game switching on and off collectively; in a logistic map a single growth rate sets both the speed of expansion and the damping of oscillations, so no choice of $r$ can match both. The simulation keeps a minority of cooperators, around $40\%$, because strategies mutate and because outside the game areas cooperation costs nothing; the equations have neither mutation nor refuge, so cooperators vanish. Finally, in the simulation the peak arrives later as the donation rate increases, which the equations do not reproduce; there the game's threshold depends on the participants' resources and fertility is capped, neither of which has a counterpart here.

\begin{table}[H]
\centering
\caption{Macroscopic features of the agent-based model, and whether Eqs.~\eqref{At_eq}--\eqref{Kdyn} reproduce them.}
\label{tab:correspondence}
\begin{tabular}{p{0.42\textwidth} p{0.36\textwidth} c}
\toprule
Feature of the simulation & Origin in the equations & Reproduced \\
\midrule
Boom, damped oscillation, stabilization & logistic map with $2 < r < 3$ & yes \\
PGG raises the peak population & capacity grows while $R_t > \theta$ & yes \\
Equilibrium independent of the PGG & capacity returns to $K_{min}$ & yes \\
Difference significant only in the transient & trajectories coincide once $K_t = K_{min}$ & yes \\
Peak rises with the donation rate & larger $\Delta K$ & yes \\
Peak rises as degradation slows & smaller $\delta_K$ & yes \\
Cooperation collapses without enforcement & $r_B > r_A$ & yes \\
Cooperators persist at $\sim 40\%$ & no mutation, no spatial refuge & no \\
Peak time shifts with the donation rate &  threshold on $R_t$ only & no \\
\bottomrule
\end{tabular}
\end{table}

\section{Analytical evaluation}
\label{sec:analytical}

    In order to shed light on the inner mechanisms of the dynamics observed numerically, we try here to get an analytical justification of the results presented in the previous Sectiom. In particular, our aim is to show theoretically the existence of the absolute maximum of $A_t$ (the number of cooperators) for a finite time $t^*$. In order to do that, we initially transform the discrete-time map given by Eq.~\eqref{At_eq} into a continuous time differential equation, so that the quantity $A$ becomes a real function of the (non-negative) real variable $t$, $A(t)$. For the sake of simplicity, we define the continuous time derivative in the most unrefined way:

\begin{equation}
    \frac{dA(t)}{dt}=\dot A(t) \equiv A_{t+1}-A_t \ .
    \label{discr_der}
\end{equation}

\noindent As a consequence, from ~\eqref{At_eq} it follows:

\begin{equation}
    \dot A(t) = (r_A-1)A(t)  - r_A\frac{A(t) + B(t)}{K(t)}A(t) \ ,
    \label{At_cont}
\end{equation}

\noindent where also $B(t), \ K(t)$ are now functions of continuous time, and we removed $\max$ because we are interested in stages of the dynamics where $A(t)$ is strictly positive. Of course, since we adopted such a big approximation, we do not expect that Eq.~\eqref{At_cont} can give back the exact, discrete dynamics (the rapid oscillations shown in Fig.~\ref{fig:comparison} will not be returned by the continuous equation), but the existence of an absolute maximum can be found, as we are going to see in the following.

The general behavior in time of $A(t)$ is naturally given by its derivative $\dot A(t)$:
it will be an increasing function when $\dot A(t)>0$ and a decreasing one when $\dot A(t)<0$. Therefore, we have to study the sign of Eq.~\eqref{At_cont}:

\begin{equation}
    (r_A-1)A(t)  - r_A\frac{A(t) + B(t)}{K(t)}A(t)  \geq 0
    \label{sign_A}
\end{equation}

\noindent which is equivalent to

\begin{equation}
   \frac{N(t)}{K(t)}  \leq 1-\frac{1}{r_A}
    \label{sign_Ab}
\end{equation}

\noindent (where $N(t)=A(t) + B(t)$). Of course, we only consider the case with $A(t)>0$ (otherwise nothing happens) and $r_A>1$ (otherwise it would be $\dot A(t)<0$ and the whole population vanishes rapidly). At the beginning of the dynamics, where we can easily assume $K(t)\simeq K_{min}$, $N(t)$ is very small, so the relation of Eq.~\eqref{sign_Ab} is satisfied, meaning that $A(t)$ increases in time. Increasing $A(t)$ also increases the overall population, until it reaches a given value $N^*$ such that the inequality in Eq.~\eqref{sign_Ab} is no longer satisfied. At this point, $A(t)$ has reached its maximum and from now on its derivative will remain negative, implying a steady decrease towards its final value, as shown in Fig.\ref{fig:timeseries}. It is worth to notice that the more K(t) increases due to the gain in fertility, the later $A(t)$ will reach its maximum, the larger will be the population itself.

\section{Discussion}
\label{sec:discussion}
The alignment between the logistic equations and the Aksum simulation highlights a fundamental property of social-ecological systems: spontaneous cooperation in the absence of centralized institutions is inherently fragile. It also says something about the simulation itself. Its macroscopic behavior is carried by three quantities: the proportion of cooperators, which declines at a rate set by the cost of cooperation; the carrying capacity, which rises while that proportion is high enough and decays otherwise; and the density feedback that ties the population to the capacity. The spatial, cognitive and metabolic machinery of the agents sets the texture of the dynamics, such as how quickly the landscape fills and how large the oscillations are, but not its structure.

In the simulation, it is observed that clusters of cooperators eventually disappear, leaving defectors to dominate a lower-yield environment. Our mathematical model explains this as the inevitable result of the cooperation ratio dropping below $\theta$. Once the ``environmental engine'' (cooperative enhancement) is turned off, the system reverts to the carrying capacity determined purely by nature ($K_{min}$), explaining the lower equilibrium population observed in non-cooperative scenarios, and explaining why that equilibrium is the same whether or not cooperation ever took place. Two further consequences of the equations, discussed in Section~\ref{sec:numerics}, are worth restating. The environment determines how large the cooperative boom becomes, but not how long it lasts: its duration is set by the cost of cooperation and the initial mix alone. And where the growth rates straddle the point at which the logistic map begins to oscillate, cooperation keeps the population steady as well as large, since the slower breeding of cooperators is what holds the population's overall growth rate below that point.

This suggests that for complex societies to sustain populations above the natural carrying capacity ($K > K_{min}$), they require mechanisms to artificially maintain $R_t > \theta$, such as the coercive authorities or ``ritual leaders'' hypothesized in the historical analysis of Axum. Mathematically, this stability requires decoupling the growth rate of defectors from that of cooperators, effectively imposing a penalty such that $r_A > r_B$ whenever resources are at risk. In social terms, this implies the emergence of institutions for ``altruistic punishment''~\citep{fehr2002} or redistribution mechanisms that exclude free-riders from the cooperative bonus; models of feedback-evolving games show that such institutions can indeed turn a transient boom into a sustained one~\citep{chen2018}. A significant challenge for future research is modeling these institutions endogenously. Rather than hard-coding a ``punishment parameter,'' future models should explore how such regulatory behaviors can emerge spontaneously from agent interactions, thereby avoiding computational artifacts and better simulating the organic rise of hierarchy in pre-state societies.

Two limitations should be stated plainly. The first concerns Aksum itself. The mechanism modeled here, in which the environmental gains of cooperation are lost once cooperation fails, is the one proposed in the geo-archaeological interpretation of Aksum's decline by~\citet{butzer1981}. That interpretation is contested: the same author later argued that societal factors generally outweigh environmental ones in historical collapse~\citep{butzer2012}; recent palaeobotanical work in Tigray finds no abrupt environmental change during the Aksumite period~\citep{ruizgiralt2021}; and the decline has been attributed instead to the loss of Red Sea trade and political fragmentation~\citep{lusini2022}. Our equations do not adjudicate between these accounts. What they provide is the dynamical signature that the environmental hypothesis would leave, a boom whose height depends on the environment and whose duration does not, against which such evidence could in future be tested. The second limitation is the list of features the equations do not reproduce, Table~\ref{tab:correspondence}. Each corresponds to a rule of the simulation that was left out, and each indicates what the next term in a slightly richer model would have to be.

\section{Conclusion}
\label{sec:conclusion}
We have presented a dynamic logistic model that captures the interplay between cooperative behavior and environmental capacity, and we have shown that it reproduces the qualitative behavior of a considerably more complicated agent-based simulation of the Aksum civilization: the boom, the damped oscillations and the stabilization; the higher peak when the cooperative mechanism is active; and, above all, the return of the population to the same final level whether or not cooperation took place. In the equations that return is not a coincidence of parameters but a consequence of their form. The model shows that environmental sustainability is contingent on maintaining a critical mass of cooperators, and that without mechanisms to maintain it, the environmental gains of cooperation are as transient as cooperation itself.

\section*{Data availability statement}
The script that generates all figures from Eqs.~\eqref{At_eq}--\eqref{Kdyn}, together with the parameter values of Table~\ref{tab:params}, is available at [repository DOI]. 

\section*{Author contributions}

\section*{Funding}

\section*{Conflict of interest}
The authors declare that the research was conducted in the absence of any commercial or financial relationships that could be construed as a potential conflict of interest.


\begin{thebibliography}{99}

\bibitem[Butzer(1981)]{butzer1981}
Butzer, K. W. (1981).
\newblock Rise and fall of Axum, Ethiopia: a geo-archaeological interpretation.
\newblock \emph{American Antiquity}, 46(3), 471--495. doi:10.2307/280596

\bibitem[Butzer(2012)]{butzer2012}
Butzer, K. W. (2012).
\newblock Collapse, environment, and society.
\newblock \emph{Proceedings of the National Academy of Sciences}, 109(10), 3632--3639. doi:10.1073/pnas.1114845109

\bibitem[Chen and Szolnoki(2018)]{chen2018}
Chen, X., \& Szolnoki, A. (2018).
\newblock Punishment and inspection for governing the commons in a feedback-evolving game.
\newblock \emph{PLOS Computational Biology}, 14(7), e1006347. doi:10.1371/journal.pcbi.1006347

\bibitem[Epstein and Axtell(1996)]{epstein1996}
Epstein, J. M., \& Axtell, R. L. (1996).
\newblock \emph{Growing Artificial Societies: Social Science from the Bottom Up}.
\newblock Brookings Institution Press / MIT Press.

\bibitem[Fehr and G\"achter(2002)]{fehr2002}
Fehr, E., \& G\"achter, S. (2002).
\newblock Altruistic punishment in humans.
\newblock \emph{Nature}, 415(6868), 137--140. doi:10.1038/415137a

\bibitem[Hardin(1968)]{hardin1968}
Hardin, G. (1968).
\newblock The Tragedy of the Commons.
\newblock \emph{Science}, 162(3859), 1243--1248. doi:10.1126/science.162.3859.1243

\bibitem[Hassell(1975)]{hassell1975}
Hassell, M. P. (1975).
\newblock Density-dependence in single-species populations.
\newblock \emph{Journal of Animal Ecology}, 44(1), 283--295.

\bibitem[Hauert et al.(2006)]{hauert2006}
Hauert, C., Holmes, M., \& Doebeli, M. (2006).
\newblock Evolutionary games and population dynamics: maintenance of cooperation in public goods games.
\newblock \emph{Proceedings of the Royal Society B}, 273(1600), 2565--2571. doi:10.1098/rspb.2006.3600

\bibitem[Krakauer et al.(2009)]{krakauer2009}
Krakauer, D. C., Page, K. M., \& Erwin, D. H. (2009).
\newblock Diversity, dilemmas, and monopolies of niche construction.
\newblock \emph{The American Naturalist}, 173(1), 26--40. doi:10.1086/593707

\bibitem[Lusini(2022)]{lusini2022}
Lusini, G. (2022).
\newblock The decline and collapse of the Kingdom of Aksum (6th--7th CE): an environmental disaster or the end of a political process?
\newblock In M. Gehler et al. (Eds.), \emph{The End of Empires} (pp. 321--336). Wiesbaden: Springer VS.

\bibitem[May(1976)]{may1976}
May, R. M. (1976).
\newblock Simple mathematical models with very complicated dynamics.
\newblock \emph{Nature}, 261(5560), 459--467. doi:10.1038/261459a0

\bibitem[Melbinger et al.(2010)]{melbinger2010}
Melbinger, A., Cremer, J., \& Frey, E. (2010).
\newblock Evolutionary game theory in growing populations.
\newblock \emph{Physical Review Letters}, 105(17), 178101. doi:10.1103/PhysRevLett.105.178101

\bibitem[Nicholson(1954)]{nicholson1954}
Nicholson, A. J. (1954).
\newblock An outline of the dynamics of animal populations.
\newblock \emph{Australian Journal of Zoology}, 2(1), 9--65. doi:10.1071/ZO9540009

\bibitem[Nowak(2006)]{nowak2006}
Nowak, M. A. (2006).
\newblock Five rules for the evolution of cooperation.
\newblock \emph{Science}, 314(5805), 1560--1563. doi:10.1126/science.1133755

\bibitem[Odling-Smee et al.(2003)]{odlingsmee2003}
Odling-Smee, F. J., Laland, K. N., \& Feldman, M. W. (2003).
\newblock \emph{Niche Construction: The Neglected Process in Evolution}.
\newblock Princeton University Press.

\bibitem[Ostrom(1990)]{ostrom1990}
Ostrom, E. (1990).
\newblock \emph{Governing the Commons: The Evolution of Institutions for Collective Action}.
\newblock Cambridge University Press.

\bibitem[Ricker(1954)]{ricker1954}
Ricker, W. E. (1954).
\newblock Stock and recruitment.
\newblock \emph{Journal of the Fisheries Research Board of Canada}, 11(5), 559--623. doi:10.1139/f54-039

\bibitem[Ruiz-Giralt et al.(2021)]{ruizgiralt2021}
Ruiz-Giralt, A., Bouchaud, C., Salavert, A., Lancelotti, C., \& D'Andrea, A. C. (2021).
\newblock Human-woodland interactions during the Pre-Aksumite and Aksumite periods in northeastern Tigray, Ethiopia: insights from the wood charcoal analyses from Mezber and Ona Adi.
\newblock \emph{Vegetation History and Archaeobotany}, 30(6), 713--728. doi:10.1007/s00334-021-00825-2

\bibitem[Santos and Pacheco(2011)]{santos2011}
Santos, F. C., \& Pacheco, J. M. (2011).
\newblock Risk of collective failure provides an escape from the tragedy of the commons.
\newblock \emph{Proceedings of the National Academy of Sciences}, 108(26), 10421--10425. doi:10.1073/pnas.1015648108

\bibitem[Tilman et al.(2020)]{tilman2020}
Tilman, A. R., Plotkin, J. B., \& Ak\c{c}ay, E. (2020).
\newblock Evolutionary games with environmental feedbacks.
\newblock \emph{Nature Communications}, 11, 915. doi:10.1038/s41467-020-14531-6

\bibitem[Vasellini et al.(2025)]{vasellini2025}
Vasellini, R., Ferrandino, G., Sernicola, L., Vilone, D., \& Mocenni, C. (2025).
\newblock Cooperative behavior in pre-state societies: an agent based approach to the Axum civilization.
\newblock \emph{Frontiers in Complex Systems}, 3, 1590952. doi:10.3389/fcpxs.2025.1590952

\bibitem[Weitz et al.(2016)]{weitz2016}
Weitz, J. S., Eksin, C., Paarporn, K., Brown, S. P., \& Ratcliff, W. C. (2016).
\newblock An oscillating tragedy of the commons in replicator dynamics with game-environment feedback.
\newblock \emph{Proceedings of the National Academy of Sciences}, 113(47), E7518--E7525. doi:10.1073/pnas.1604096113

\end{thebibliography}
\end{document}